\documentstyle[11pt,epsfig]{article}
\title{\bf Role of Chaplygin gas as  geometrical dark energy in anisotropic brane gravity}
\author{Malihe Heydari-Fard\thanks{email:
m.heydarifard@mail.sbu.ac.ir} \hspace{0.5mm} and Hamid R.
Sepangi\thanks{email: hr-sepangi@sbu.ac.ir}
\\ {\small Department of Physics, Shahid Beheshti University, Evin, Tehran 19839, Iran}}
\begin{document}
\maketitle 
\begin{abstract}
We consider an anisotropic brane-world model with Bianchi type I
and V geometry, without mirror symmetry or any form of junction
conditions. The generalized Chaplygin gas, which interpolates
between a high density relativistic era and a non-relativistic
matter phase, is a popular candidate for the present accelerated
expansion of the universe. Considering the generalized Chaplygin
gas as a geometrical dark energy, we obtain the general solutions
in an exact parametric form for both Bianchi type I and V
space-times. Finally, we study the behavior of the observationally
important parameters such as the shear, anisotropic and
deceleration parameter in this model.
\vspace{2mm}\noindent\\
PACS numbers: 04.50.-h, 04.20.-q, 04.20.Jb
\vspace{5mm}\\
\end{abstract}
\section{Introduction}
From a large number of observational evidence, the observable
universe is presently undergoing an accelerated expansion. As an
alternative to both the cosmological constant and quintessence, it
is also possible to explain the acceleration of the universe by
introducing a cosmic fluid component with an exotic equation of
state, called Chaplygin gas. The Chaplygin gas model describes a
transition from a universe filled with dust-like matter to an
accelerated expanding stage. The Generalized Chaplygin Gas (GCG)
model, introduced in \cite{3} and elaborated in \cite{4}, is
described by a perfect fluid obeying an exotic equation of state
\begin{eqnarray}
p_{ch}=-\frac{A}{\rho_{ch}^{\,\,\,\,\alpha}},
\end{eqnarray}
where $A$ is a positive constant and $0<\alpha\leq1$. The original
Chaplygin gas corresponds to $\alpha=1$. However, recent analyses
based on the latest of type-Ia Supernovae data have yielded rather
surprising results, namely that $\alpha>1$ and that  there is a
degeneracy between the GCG  and  xCDM models in the form of a
phantom-like energy component \cite{Bertolami1}. An attractive
feature of the model is that it can naturally explain both dark
energy and dark matter \cite{5}. Within the framework of
Friedmann-Robertson-Walker cosmology, this equation of state
leads, after having inserted into the relativistic energy
conservation equation, to an energy density evolving as
\begin{eqnarray}
\rho_{ch}=\left[A+\frac{C}{a^{3(1+\alpha)}}\right]^{\frac{1}{1+\alpha}},
\end{eqnarray}
where $a$ is the scale-factor of the universe and $C$ is an
integration constant which should be positive for a well-defined
$\rho_{ch}$ at all times. Hence, we see that $\rho_{ch}\sim
a^{-3}$ at early times, that is, $\rho_{ch}$ behaves as matter
while at late times it behaves like a cosmological constant
$\rho_{ch} \sim$ constant. The Chaplygin gas appears in the
stabilization of branes in Schwarzschild anti-de Sitter (AdS)
black hole bulks as a critical theory at the horizon \cite{Kame}
and in the string analysis of black holes in three dimensions
\cite{Kama}. The Chaplygin gas also appears as an effective fluid
associated with $d$-branes \cite{6} and can also be derived from
the Born-Infeld action \cite{9}. An interesting range of models
have been found to be consistent with the SNe Ia data \cite{SNe},
CMB experiments \cite{CMB} and other observational data
\cite{other}. The cosmological implications of the Chaplygin gas
model have been intensively investigated in the literature
\cite{Mak,c1,C1}.

The idea that our familiar 4-dimensional ($4D$) space-time is a
hypersurface (brane) in a 5D space-time (bulk)
\cite{Nima,Randall,Dvali} has been under detailed elaboration
during the last decade. According to this brane-world scenario,
all matter and gauge interactions reside on the brane, while
gravity can propagate in the 5D space-time. Several brane-world
cosmologies have been proposed in the context of the
Randall-Sundrum (RS) formulations \cite{Randall}, defined in a
5-dimensional anti-de Sitter space-time. The dynamics of these
models feature boundary terms in the action and sometimes mirror
symmetry, such that bulk gravitational waves interfere with the
brane-world motion. This usually comes together with junction
conditions producing an algebraic relationship between the
extrinsic curvature and the confined matter \cite{israel,Battye}.
The consequence is that the Friedman equation acquires an
additional term which is proportional to the square of energy
density of the confined matter field \cite{Cline,Binetruy}. This
term was initially considered as a possible solution to the
accelerated expansion of the universe. However, soon it was
realized to be incompatible with the big bang nucleosynthesis,
requiring additional fixes \cite{Binetruy}.

Brane-world scenarios under more general conditions and still
compatible with the brane-world program have also been rather
extensively studied over the past decade where it has been shown
that it is possible to find a richer set of cosmological solutions
in accordance with the current observations \cite{Maia}. Under
these conditions, without using $Z_2$ symmetry or without
postulating any junction condition, Friedman equation is modified
by a geometrical term which is defined in terms of the extrinsic
curvature, leading to a geometrical interpretation for dark energy
\cite{maia}. There has also been arguments concerning the
uniqueness of the junction conditions. Indeed, other forms of
junction conditions exist, so that different conditions may lead
to different physical results \cite{Battye}. Furthermore, these
conditions cannot be used when more than one non-compact extra
dimension is involved. Against this background, an interesting
higher-dimensional model was introduced in \cite{Rubakov} where
particles are trapped on a 4D hypersurface by the action of a
confining potential. The dynamics of test particles confined to a
brane by the action of such potential at the classical and quantum
levels were studied in \cite{shahram}. In \cite{fard}, the same
brane-world model was studied, offering a geometrical explanation
for the accelerated expansion of the universe. A geometrical
explanation for the generalized Chaplygin gas was considered in
\cite{gas} along the same line. In this paper, we consider an
anisotropic brane-world model with Bianchi type I and V geometries
filled with a geometrical Chaplygin gas. The behavior of the
observationally important physical quantities is studied in this
scenario.
\section{The model}
The embedding of the brane-world in the bulk plays an essential
role in the covariant formulation of the brane-world gravity,
because it tells us how the Einstein-Hilbert dynamics of the bulk
is transferred to the brane. However, there are many different
ways to embed a manifold into another, classified as local,
global, isometric, conformal, rigid, deformable, analytic or
differentiable. The choice of one or other depends on what the
embedded manifold is supposed to do.

Let us present a brief review of the model proposed in
\cite{maia}. Consider the background manifold $ \overline{V}_{4} $
isometrically embedded in a pseudo-Riemannian manifold $ V_{m}$ by
the map ${ \cal Y}: \overline{V}_{4}\rightarrow  V_{m} $, with
$m=4+n$ components ${\cal Y}^{A}$ such that
\begin{eqnarray}\label{a1}
{\cal G} _{AB} {\cal Y}^{A}_{,\mu } {\cal Y}^{B}_{,\nu}=
\bar{g}_{\mu \nu}  , \hspace{.5 cm} {\cal G}_{AB}{\cal
Y}^{A}_{,\mu}\bar{{\cal N}}^{B}_{a} = 0  ,\hspace{.5 cm}  {\cal
G}_{AB}\bar{{\cal N}}^{A}_{a}\bar{{\cal N}}^{B}_{b} =\bar{g}_{ab},
\end{eqnarray}
where $ {\cal G}_{AB} $  $ ( \bar{g}_{\mu\nu} )$ is the metric of
the bulk (brane) space  $  V_{m}  (\overline{V}_{4}) $ in
arbitrary coordinates, $ \{ {\cal Y}^{A} \} $   $  (\{ x^{\mu} \})
$  is the  basis of the bulk (brane) and  ${\cal N}^{A}_{a}$ are
$n$ normal unit vectors orthogonal to the brane. According to Nash
\cite{Nash}, we may continuously perturb $\overline{V}_{4}$ along
a normal direction in the bulk, to obtain another submanifold of
the same bulk, provided the embedding functions remain regular.
Perturbation of $\bar{V}_{4}$ in a sufficiently small neighborhood
of the brane along an arbitrary transverse direction $\xi$ is
given by
\begin{eqnarray}\label{a2}
{\cal Z}^{A} = {\cal Y}^{A} + ({\cal L}_{\xi}{\cal Y})^{A},
\hspace{.5 cm} {\cal N}^{A} = \bar{{\cal N}}^{A} + ({\cal
L}_{\xi}\bar{{\cal N}})^{A}=\bar{{\cal N}}^{A},
\end{eqnarray}
where $\cal L$ represents the Lie derivative and $\xi^{a}$ $(a =
1,2,...,n)$ is a small parameter along ${\cal N}^{A}_{a}$,
parameterizing the extra noncompact dimensions. By choosing $\xi$
orthogonal to the brane, we ensure gauge independency and have
perturbations of the embedding along a single orthogonal extra
direction $\bar{{\cal N}}_{a}$ giving local coordinates of the
perturbed brane as
\begin{eqnarray}\label{a2}
{\cal Z}^{A}_{,\mu} = {\cal Y}^{A}_{,\mu} + \xi^{a}\bar{{\cal
N}}^{A}_{a,\mu}.
\end{eqnarray}
The above  assumptions lead to the embedding equations of the
perturbed geometry
\begin{eqnarray}\label{a3}
{\cal G}_{AB}{\cal Z}_{,\mu }^{A}{\cal Z}_{,\nu }^{B}=g_{\mu \nu
},\hspace{0.5cm}{\cal G}_{AB}{\cal Z}_{,\mu
}^{A}{\cal N}_{a}^{B}=g_{\mu a},\hspace{0.5cm}{\cal G}_{AB}{\cal N}_{a}^{A}%
{\cal N}_{b}^{B}={g}_{ab}.
\end{eqnarray}
From these equations it follows that
\begin{eqnarray}\label{a4}
g^{\mu\nu}{\cal Z}_{,\mu }^{A}{\cal Z}_{,\nu }^{B}={\cal
G}^{AB}-g^{ab}{\cal N}_{a}^{A} {\cal N}_{b}^{B},
\end{eqnarray}
and also the components of the perturbed geometry
\begin{eqnarray}\label{a5}
g_{\mu\nu}={\cal G}_{AB}{\cal Z}_{,\mu }^{A}{\cal Z}_{,\nu
}^{B}=\bar{g}_{\mu\nu}-2\xi^{a}\bar{K}_{\mu\nu
a}+\xi^{a}\xi^{b}\left[\bar{g}^{\alpha\beta}\bar{K}_{\mu\alpha
a}\bar{K}_{\nu\beta b}+g^{cd}\bar{A}_{\mu ca}\bar{A}_{\nu
db}\right],
\end{eqnarray}
\begin{eqnarray}\label{a6}
g_{\mu b}={\cal G}_{AB}{\cal Z}_{,\mu }^{A}{\cal N}_{b
}^{B}=\xi^{a}A_{\mu ab},
\end{eqnarray}
\begin{eqnarray}\label{a7}
g_{ab}={\cal G}_{AB}{\cal N}_{a}^{A}{\cal N}_{b
}^{B}=\bar{g}_{ab},
\end{eqnarray}
\begin{eqnarray}\label{a8}
K_{\mu \nu a}=-{\cal G}_{AB}{\cal N}_{a,\mu }^{A}{\cal Z}_{,\nu
}^{B}=\bar{K}_{\mu\nu
a}-\xi^{b}\left[\bar{g}^{\alpha\beta}\bar{K}_{\mu\alpha
a}\bar{K}_{\nu\beta b}+g^{cd}\bar{A}_{\mu ca}\bar{A}_{\nu
db}\right],
\end{eqnarray}
\begin{eqnarray}\label{a9}
A_{\mu ab}={\cal G}_{AB}{\cal N}_{a,\mu }^{A}{\cal N}_{b
}^{B}=\bar{A}_{\mu ab},
\end{eqnarray}
where $A_{\mu ab}$ represents the twisting vector fields and
$\bar{K}_{\mu \nu a}$, $K_{\mu \nu a}$ represent the extrinsic
curvature of the original and perturbed brane respectively.
Comparing equations (\ref{a5}) and (\ref{a8}), we obtain
\begin{eqnarray}\label{a10}
{K}_{\mu \nu a}=-\frac{1}{2}\frac{\partial {g}_{\mu \nu }}{%
\partial \xi ^{a}},
\end{eqnarray}%
which is the generalized York's relation and shows how the
extrinsic curvature propagates as a result of the propagation of
the metric in the direction of extra dimensions. The components of
the Riemann tensor of the bulk written in the embedding vielbein
$\{{\cal Z}^{A}_{, \alpha}, {\cal N}^A_a \}$, lead to the
Gauss-Codazzi and Ricci equations, respectively \cite{Book}
\begin{eqnarray}\label{a11}
R_{\alpha \beta \gamma \delta}=2g^{ab}K_{\alpha[ \gamma
a}K_{\delta] \beta b}+{\cal R}_{ABCD}{\cal Z} ^{A}_{,\alpha}{\cal
Z} ^{B}_{,\beta}{\cal Z} ^{C}_{,\gamma} {\cal Z}^{D}_{,\delta},
\end{eqnarray}
\begin{eqnarray}\label{a12}
2K_{\alpha [\gamma c; \delta]}=2g^{ab}A_{[\gamma ac}K_{ \delta]
\alpha b}+{\cal R}_{ABCD}{\cal Z} ^{A}_{,\alpha} {\cal N}^{B}_{c}
{\cal Z} ^{C}_{,\gamma} {\cal Z}^{D}_{,\delta},
\end{eqnarray}
\begin{eqnarray}\label{a13}
2A_{[\gamma ab;\delta]}=-2g^{cd}A_{[\gamma
ca}A_{\delta]db}-g^{cd}K_{[\gamma ca}K_{\delta]db}-{\cal
R}_{ABCD}{\cal N}^{A}_{a}{\cal N}^{B}_{b}{\cal
Z}^{C}_{,\gamma}{\cal Z}^{D}_{,\delta},
\end{eqnarray}
where ${\cal R}_{ABCD}$ and $R_{\alpha\beta\gamma\delta}$ are the
Riemann tensors for the bulk and the perturbed brane respectively.
Contracting the Gauss equation (\ref{a11}) on ${\alpha}$ and
${\gamma}$ we find
\begin{eqnarray}\label{a14}
R_{\mu\nu}=(K_{\mu\alpha c}K_{\nu}^{\,\,\,\,\alpha c}-K_{c} K_{\mu
\nu }^{\,\,\,\ c})+{\cal R}_{AB} {\cal Z}^{A}_{,\mu} {\cal
Z}^{B}_{,\nu}-g^{ab}{\cal R}_{ABCD}{\cal N}^{A}_{a}{\cal
Z}^{B}_{,\mu}{\cal Z}^{C}_{,\nu}{\cal N}^{D}_{b}.
\end{eqnarray}
A further contraction then gives the Ricci scalar
\begin{eqnarray}\label{a15}
R=(K_{\mu\nu a}K^{\mu\nu a}-K_{a} K^{a})+{\cal R}-2g^{ab}{\cal
R}_{AB}{\cal N}^{A}_{a}{\cal N}^{B}_{b}.
\end{eqnarray}
Therefore, the Einstein-Hilbert action for the bulk geometry in
$m$-dimensions can be written as
\begin{eqnarray}
S&=&\frac{1}{2\alpha_{*}}\int({\cal R}-2\Lambda^{(b)})\sqrt{{\cal
G}}d^mx\nonumber\\
&=&\frac{1}{2\alpha_{*}}\int\left[R-(K_{\mu\nu a}K^{\mu\nu
a}-K_{a} K^{a})+2g^{ab}{\cal R}_{AB}{\cal N}^{A}_{a}{\cal
N}^{B}_{b}-2\Lambda^{(b)}\right]\sqrt{{\cal G}}d^mx.
\end{eqnarray}
Variation of the action with respect to ${\cal G}_{AB}$ gives the
Einstein field equations in the bulk
\begin{eqnarray}\label{a16}
{\cal R}_{AB}-\frac{1}{2}{\cal R}{\cal G}_{AB}=\alpha_{*}
T^{*}_{AB}-\Lambda^{(b)} {\cal G}_{AB},
\end{eqnarray}
where $\alpha_{*}=\frac{1}{M_{*}^{m-2}}$ and $\Lambda^{(b)}$ is
the cosmological constant of the bulk space. The vielbein
components of the energy-momentum tensor are given by
\begin{eqnarray}\label{a17}
T_{\mu\nu}^{*}=T_{AB}^{*}{\cal Z}^{A}_{,\mu}{\cal
Z}^{B}_{,\nu},\hspace{.5 cm} T_{\mu a}^{*}=T_{AB}^{*}{\cal
Z}^{A}_{,\mu}{\cal N}^{B}_{a},\hspace{.5 cm}
T_{ab}^{*}=T_{AB}^{*}{\cal N}^{A}_{a}{\cal N}^{B}_{b}.
\end{eqnarray}
The tangent components follow from the contractions of equation
$(\ref{a16})$ with ${\cal Z}^{A}_{,\mu}{\cal Z}^{B}_{,\nu}$. After
using equations (\ref{a14}) and (\ref{a15}) we obtain the
``gravi-tensor'' equation
\begin{eqnarray}\label{a18}
R_{\mu\nu}-\frac{1}{2}Rg_{\mu\nu}-Q_{\mu\nu}+g^{ad}{\cal
R}_{ABCD}{\cal N}^{A}_{a}{\cal Z}^{B}_{,\mu}{\cal
Z}^{C}_{,\nu}{\cal N}^{D}_{d}-g^{ab}{\cal R}_{AB}{\cal
N}^{A}_{a}{\cal
N}^{B}_{b}g_{\mu\nu}=\alpha_{*}T_{\mu\nu}^{*}-\Lambda^{(b)}g_{\mu\nu},
\end{eqnarray}
where
\begin{eqnarray}\label{a19}
Q_{\mu\nu}=g^{ab}\left(K^{\rho}_{\mu a}K_{\rho\nu b}-K_aK_{\mu\nu
b}\right)-\frac{1}{2} \left(K_{\alpha\beta a}K^{\alpha\beta
a}-K_aK^a\right)g_{\mu\nu}.
\end{eqnarray}
By a direct calculation we can see that the extra term
$Q_{\mu\nu}$ is an independently conserved quantity. On the other
hand, the trace of the Codazzi equation gives the ``gravi-vector''
equation
\begin{eqnarray}\label{a20}
K^{\rho}_{\mu a;\rho}&-&(g^{\mu\nu}K_{\mu\nu a})_{,\mu}-(A_{\rho
ca}K^{\rho c}_{\mu}-A_{\mu
ca}g^{\alpha\beta}K^{c}_{\alpha\beta})+2g^{cd}{\cal R}_{ABCD}{\cal
N}^{A}_{a}{\cal N}^{B}_{c}{\cal Z}^{C}_{,\mu}{\cal
N}^{D}_{d}\nonumber\\
 &=&\alpha_{*}\left(T_{\mu
a}^{*}-\frac{1}{n+2}T^{*}g_{\mu
a}\right)+\frac{2}{n+2}\Lambda^{(b)}g_{\mu a}.
\end{eqnarray}
Finally, the ``gravi-scalar'' equation is obtained from equations
(\ref{a15}) and (\ref{a16})
\begin{eqnarray}\label{a21}
R-K_{\mu\nu a}K^{\mu\nu
a}+K_{a}K^{a}=-2\alpha_{*}\left(g^{ab}T_{ab}^{*}-\frac{n-1}{n+2}T^{*}\right)+2\Lambda^{(b)}\left[
n-\frac{(n-1)(n+4)}{(n+2)}\right].
\end{eqnarray}
In its most general form, without assuming extra dimensional
matter, the confinement hypothesis states that the only
non-vanishing components of $T_{AB}$ are the tangent components
$T_{\mu\nu}$ representing the confined sources. Therefore we set
\begin{eqnarray}\label{a22}
\alpha_{*}T_{\mu\nu}^{*}=8\pi GT_{\mu\nu},\hspace{.5 cm}
\alpha_{*}T_{\mu a}^{*}=0,\hspace{.5 cm} \alpha_{*}T_{ab}^{*}=0.
\end{eqnarray}
Equations  (\ref{a18})-(\ref{a21}) represent the most general
equations of motion of a brane-world, compatible with the
differentiable embedding in a $m$-dimensional bulk defined by the
Einstein equations. Clearly, the usual Einstein equations are
recovered when all elements of the extrinsic geometry are removed
from those equations.

The geometrical approach considered here is based on three basic
postulates, namely, the confinement of the standard gauge
interactions to the brane, the existence of quantum gravity in the
bulk and finally, the embedding of the brane-world. All other
model dependent properties such as warped metric, mirror
symmetries, radion or extra scalar fields, fine tuning parameters
like the tension of the brane and the choice of a junction
condition are left out as much as possible in our calculations
\cite{Maia}.
\section{Field equations and observational parameters in anisotropic brane}
In what follows, we will investigate the influence of the
extrinsic curvature terms on the anisotropic universe described by
Bianchi type I and V geometries. From a formal point of view these
two geometries are described by the line element
\begin{eqnarray}
ds^2 = - dt^2 + a_1^2(t) dx^2 + a_2^2(t) e^{-2\beta x} dy^2 +
a_3^2(t) e^{-2\beta x} dz^2,\label{7}
\end{eqnarray}
where $a_i(t), i=1,2,3$ are the expansion factors in different
spatial directions. The  metric for the Bianchi type I geometry
corresponds to the case $\beta=0$, while for the Bianchi type V we
have $\beta=1$. Let us define the following variables \cite{Chen}
\begin{eqnarray}
v=\prod_{i=1}^3 a_i,\hspace{.5 cm} H_i=\frac{\dot
a_i}{a_i},i=1,2,3,\hspace{.5 cm} 3H=\sum_{i=1}^3 H_i,\hspace{.5
cm} \Delta H_i &=& H_i - H,i=1,2,3.\label{8}
\end{eqnarray}
In above equations, $v$ is the volume scale factor, $H_{i},
i=1,2,3$ are the directional Hubble parameters, and $H$ is the
mean Hubble parameter. The physical quantities of observational
importance in cosmology are the expansion scalar $\Theta$, the
mean anisotropy parameter $A$, the shear scalar parameter
$\sigma^2$, and the deceleration parameter $q$, which are defined
according to
\begin{eqnarray}
\Theta=3H,\hspace{.5 cm}3A=\sum_{i=1}^3 \left( \frac{\Delta
H_i}{H} \right)^2,\hspace{.5
cm}\sigma^2=\frac{1}{2}\sigma_{ij}\sigma^{ij}=\frac{1}{2}\sum_{i=1}^3
H_i^2-3H^2,\hspace{.5 cm}q=\frac{d}{d t}\left(\frac{1}{H}\right)-1
.\label{9}
\end{eqnarray}
The sign of the deceleration parameter indicates how the universe
expands. A positive sign for $q$ corresponds to the standard
decelerating models whereas a negative sign indicates an
accelerating expansion at late times. We note that $A=0$ for an
isotropic expansion.

Let us assume that the confined source on the brane is a perfect
fluid with a linear barotropic equation of state, namely
$p=(\gamma-1)\rho$ with $1\leq\gamma\leq2$. In this paper we
restrict our analysis to a five-dimensional bulk with a constant
curvature characterized by the Riemann tensor
\begin{eqnarray}
{\cal R}_{ABCD}=k_{*}({\cal G}_{AC}{\cal G}_{BD}-{\cal
G}_{AD}{\cal G}_{BC}),\label{1}
\end{eqnarray}
where $k_*$ denotes the bulk constant curvature. In the flat case
$k_*=0$ and in the de Sitter and anti-de Sitter cases we may write
$k_*=\pm\frac{\Lambda^{(b)}}{6}$ respectively. Assuming $g_{55}=1$
and using the Gauss-Codazzi equations, we obtain
\begin{eqnarray}
R_{\alpha\beta\gamma\delta} =
(K_{\alpha\gamma}K_{\beta\delta}-K_{\alpha\delta}K_{\beta\gamma})
+ k_{*}
(g_{\alpha\gamma}g_{\beta\delta}-g_{\alpha\delta}g_{\beta\gamma}),\label{22}
\end{eqnarray}
\begin{eqnarray}
K_{\alpha[\beta;\gamma]} = 0.\label{3}
\end{eqnarray}
The equations of motion derived in the previous section can be
obtained directly from equations (\ref{22}) and (\ref{3}). The
result is Einstein equations as modified by the presence of the
extrinsic curvature
\begin{eqnarray}
G_{\mu\nu} = 8\pi G T_{\mu\nu}-\Lambda g_{\mu\nu} +
Q_{\mu\nu},\label{4}
\end{eqnarray}
where $\Lambda=-3k_{*}+\Lambda^{(b)}$ is the effective
cosmological constant in four dimensions with $Q_{\mu\nu}$ being a
completely geometrical quantity given by
\begin{eqnarray}
Q_{\mu\nu}=\left(KK_{\mu\nu}- K_{\mu\alpha
}K^{\alpha}_{\nu}\right)+\frac{1}{2}
\left(K_{\alpha\beta}K^{\alpha\beta}-K^2\right)g_{\mu\nu},\label{5}
\end{eqnarray}
where $K=g^{\mu\nu}K_{\mu\nu}$. Using the York relation
\begin{eqnarray}
K_{\mu \nu a}=-\frac{1}{2}\frac{\partial
g_{\mu\nu}}{\partial\xi^{a}},\label{10}
\end{eqnarray}
we realize that in a diagonal metric, $K_{\mu\nu a}$ is diagonal.
After separating the spatial components, the Codazzi equations
reduce to (here $\alpha,\beta,\gamma,\sigma=1,2,3$)
\begin{eqnarray}
K^{\alpha}_{\,\,\,\gamma a,\sigma}+K^{\beta}_{\,\,\,\gamma
a}\Gamma^{\alpha}_{\,\,\,\beta\sigma}= K^{\alpha}_{\,\,\,\sigma
a,\gamma}+K^{\beta}_{\,\,\,\sigma
a}\Gamma^{\alpha}_{\,\,\,\beta\gamma},\label{11}
\end{eqnarray}
\begin{eqnarray}
K^{\alpha}_{\,\,\,\gamma
a,0}+\frac{\dot{a_i}}{a_i}K^{\alpha}_{\,\,\,\gamma
a}=\frac{\dot{a_i}}{a_i}\delta^{\alpha}_{\,\,\,
\gamma}K^{0}_{\,\,\,0a},\hspace{.5 cm}i=1,2,3.\label{12}
\end{eqnarray}
The first equation gives $K^{1}_{\,\,\,1 a,\sigma}=0$  for
$\sigma\neq1$, since $K^{1}_{\,\,\,1 a}$ does not depend on the
spatial coordinates. Repeating the same procedure for
$\alpha,\gamma=i, i=2,3,$ we obtain $K^{2}_{\,\,\,2 a,\sigma}=0$
for $\sigma\neq2$ and $K^{3}_{\,\,\,3 a,\sigma}=0$ for
$\sigma\neq3$. This shows that $K^{1}_{\,\,\,1 a}$,
$K^{2}_{\,\,\,2 a}$ and $K^{3}_{\,\,\,3 a}$ are functions of $t$
and the choice $K^{1}_{\,\,\,1 a}=K^{2}_{\,\,\,2 a}=K^{3}_{\,\,\,3
a}=b_{a}(t)$, where $b_{a}(t)$ are arbitrary functions of $t$,
would simplify our analysis. Now from the second equation we
obtain
\begin{eqnarray}
\dot{b_{a}}+\frac{\dot{a}_i}{a_i}b_{a}=\frac{\dot{a}_i}{a_i}K^{0}_{\,\,\,0
a},\hspace{.5 cm}i=1,2,3.\label{S}
\end{eqnarray}
Summing equations (\ref{S}) we find
\begin{eqnarray}
K_{00
a}=-\left(\frac{3\dot{b}_{a}v}{\dot{v}}+b_{a}\right).\label{14}
\end{eqnarray}
For $\mu,\nu=1,2,3$ we obtain
\begin{eqnarray}
K_{\mu\nu a}=b_{a}g_{\mu\nu}.\label{15}
\end{eqnarray}
Note that in equation (\ref{4}) we have considered a
five-dimensional bulk ($m=5$), thus the functions $b_a$
($a=1,2,...,m-4$) reduce to only one function, namely $b_1$.
Denoting $b_{1}=b$, $\theta=\frac{\dot{b}}{b}$ and
$\Theta=\frac{\dot{v}}{v}$, we find from equation (\ref{5}) that
\begin{eqnarray}
Q_{\mu\nu}=-3 b^2\left(\frac{2\theta}{\Theta}+1\right)g_{\mu\nu},
\hspace{5mm}\mu,\nu=1,2,3,\hspace{.5 cm} Q_{00}=3 b^2.\label{100}
\end{eqnarray}
As we have noted before, $Q_{\mu\nu}$ is an independently
conserved quantity, that is $Q^{\mu\nu}_{\,\,\,\,;\nu}=0$,
suggesting an analogy with the energy momentum of an uncoupled
non-conventional energy source. We see that solution (\ref{100})
depends on the arbitrary function $b(t)$. To find the dynamical
role of this function and to compare the compatibility of such
geometrical model with the present experimental data, we assume
$Q_{\mu\nu}$ to be a conserved energy-momentum tensor and take the
GCG as an example
\begin{eqnarray}\label{11}
Q_{\mu\nu}\equiv\frac{1}{8\pi G}\left[(\rho_{ch}+p_{ch})
u_{\mu}u_{\nu}+p_{ch} g_{\mu\nu}\right] ,\hspace{.5 cm}
p_{ch}=-\frac{A}{\rho_{ch}^{\,\,\,\,\alpha}},\label{11}
\end{eqnarray}
where $A$ and $\alpha$ are positive constants. Comparing
$Q_{\mu\nu}, \mu,\nu=1,2,3$ and $Q_{00}$ from equation (\ref{11})
with the components of $Q_{\mu\nu}$ and $Q_{00}$ given by equation
(\ref{100}), we obtain
\begin{eqnarray}\
p_{ch}=-\frac{3b^2}{8\pi G}\left(\frac{2\theta}{\Theta}+1\right)
,\hspace{.5 cm}\rho_{ch}=\frac{3b^2}{8\pi G}.\label{12}
\end{eqnarray}
Use of the above equations leads to an equation for $b(t)$
\begin{eqnarray}\label{9}
\left(\frac{2v\dot{b}}{\dot{v}b}+1\right)=A\left(\frac{3b^2}{8\pi
G}\right) ^{-(1+\alpha)},\label{13}
\end{eqnarray}
for which the solution is
\begin{eqnarray}\label{14}
b(t)=\left(\frac{8\pi G}{3}\right)^{\frac{1}{2}}
\left[A+\frac{C}{v^{(1+\alpha)}}\right]^{\frac{1}{2(1+\alpha)}},\label{14}
\end{eqnarray}
where $C$ is an integration constant. Using equation (\ref{12})
and this solution, the energy density of the GCG becomes
\begin{eqnarray}\label{15}
\rho_{ch}=\rho_{ch_{0}}\left[A_{s}
+\frac{(1-A_{s})}{v^{(1+\alpha)}}\right]^{\frac{1}{(1+\alpha)}},\label{15}
\end{eqnarray}
where $\rho_{ch_{0}}$ is the GCG density at the present time,
$A_{s}=A\rho_{ch_{0}}^{-(1+\alpha)}$ is a dimensionless quantity
related to the speed of sound for the GCG today, $v_{s}^{2}=\alpha
A \rho_{ch_{0}}^{-(1+\alpha)}$ and
$C=\rho_{ch_{0}}^{(1+\alpha)}-A$. From equation (\ref{15}), we see
that for $A_{s}=0$ the GCG behaves like matter, whereas for
$A_{s}=1$ it behaves as a cosmological constant and when
$0<A_{s}<1$, the model predicts a behavior as that between a
matter phase in the past and a negative dark energy regime at late
times. This particular behavior of the GCG inspired some authors
to propose a unified scheme for the cosmological ``dark sector,''
an interesting idea which has been considered in many different
contexts.

Now, using the geometrical energy density for $Q_{\mu\nu}$, the
field equations on the anisotropic brane become
\begin{eqnarray}
3 \dot H + \sum_{i=1}^3 H_i^2=\Lambda - 4\pi
G\rho_0(3\gamma-2)v^{-\gamma}+8\pi
G\rho_{ch_{0}}\left[A_s+\frac{(1-A_s)}{v^{(1+\alpha)}}\right]
^{\frac{-\alpha}{1+\alpha}}\left[A_s-\frac{(1-A_s)}{2v^{(1+\alpha)}}\right],\label{17}
\end{eqnarray}
\begin{eqnarray}
\frac{1}{v} \frac{d}{d t} (v H)=2\beta^2v^{-2/3}+\Lambda -4\pi
G\rho_0(\gamma-2)v^{-\gamma} +8\pi
G\rho_{ch_{0}}\left[A_s+\frac{(1-A_s)}{v^{(1+\alpha)}}\right]
^{\frac{-\alpha}{1+\alpha}}\left[A_s+\frac{(1-A_s)}{2v^{(1+\alpha)}}\right].\label{18}
\end{eqnarray}
For $\beta=0$ we obtain the field equations for Bianchi type I
geometry, while $\beta=1$ gives the Bianchi type $V$ equations on
the anisotropic brane. Using the relation $H=\frac{\dot{v}}{3v}$,
we can rewrite equation (\ref{18}) in the form
\begin{eqnarray}
\ddot v=6\beta^2v^{1/3}+3 \Lambda v -12\pi G
\rho_{0}(\gamma-2)v^{1-\gamma} +24\pi G v\rho_{ch_{0}}
\left[A_s+\frac{(1-A_s)}{v^{(1+\alpha)}}\right]
^{\frac{-\alpha}{1+\alpha}}\left[A_s+\frac{(1-A_s)}{2v^{(1+\alpha)}}\right].\label{32}
\end{eqnarray}
The general solution of equation (\ref{32}) becomes
\begin{eqnarray}
t - t_0 = \int \frac{dv}{\sqrt{ 9\beta^2v^{4/3}+3\Lambda v^2+24\pi
G \rho_{0}v^{2-\gamma} +f(v)+C}},\label{33}
\end{eqnarray}
where
\begin{eqnarray}
f(v)&=&\frac{24\pi G \rho_{ch_{0}}
}{2+\alpha}v\left(1+\frac{A_sv^{1+\alpha}}{1-A_s}\right)^{\frac{\alpha}{1+\alpha}}
\left[1+
A_s\left(v^{1+\alpha}-1\right)\right]^{\frac{-\alpha}{1+\alpha}}\nonumber\\
&\times&[2A_sv^{1+\alpha}{_2F_1}\left(\frac{2+\alpha}{1+\alpha},\frac{\alpha}{1+\alpha}
\frac{3+2\alpha}{1+\alpha},\frac{A_sv^{1+\alpha}}{A_s-1}\right)\nonumber\\
&+&{(2+\alpha)(1-A_s)}
{_2F_1}\left(\frac{1}{1+\alpha},\frac{\alpha}{1+\alpha}
\frac{2+\alpha}{1+\alpha},\frac{A_sv^{1+\alpha}}{A_s-1}\right)],\label{hypergeometry}
\end{eqnarray}
and $C$ is a constant of integration. The time variation of the
physically important parameters described above in the exact
parametric form, with $v$ taken as a parameter, is given by
\begin{eqnarray}
\Theta=3H=\frac{\sqrt{9\beta^2v^{4/3}+3\Lambda v^2+24\pi G
\rho_{0}v^{2-\gamma} +f(v)+C}}{v},\label{35}
\end{eqnarray}
\begin{eqnarray}
a_i=a_{0i} v^{1/3} \exp \left[\int {\frac{ h_idv
}{v\sqrt{9\beta^2v^{4/3}+3\Lambda v^2+24\pi G \rho_{0}v^{2-\gamma}
+f(v)+C}} }\right], \quad i=1,2,3,\label{37}
\end{eqnarray}
\begin{eqnarray}
A=\frac{3h^2}{9\beta^2v^{4/3}+3\Lambda v^2+24\pi G
\rho_{0}v^{2-\gamma} +f(v)+C},\label{36}
\end{eqnarray}
\begin{eqnarray}
\sigma^2=\frac{h^2}{2v^2},\label{36}
\end{eqnarray}
\begin{eqnarray}
q=2-\frac{ 9\beta^2v^{4/3}+3\Lambda v^2+12\pi G
\rho_{0}(2-\gamma)v^{2-\gamma}+24\pi G
\rho_{ch_{0}}v^2\left[A_s+\frac{(1-A_s)}{v^{(1+\alpha)}}\right]
^{\frac{-\alpha}{1+\alpha}}\left[A_s+\frac{(1-A_s)}{2v^{(1+\alpha)}}\right]}{3
\beta^2v^{4/3}+\Lambda v^2+8\pi G \rho_{0}v^{2-\gamma}
+\frac{1}{3}f(v)+C},\label{38}
\end{eqnarray}
where $h_i, i=1,2,3$ are constants of integration and
$h^2=\sum_{i=1}^3 h_i^2$. The GCG provides an interesting
candidate for the present accelerated expansion of the universe
without resorting to an effective cosmological constant. To this
end we consider $\Lambda=0$ and show that, within the context of
the present model, the geometrical Chaplygin gas can be used to
account for the accelerated expansion of the universe. Therefore
for $\Lambda=0$ and $\alpha=1$, the dynamics of the Bianchi type I
universe is controlled by the Chaplygin gas equation of state
parameter $A_s$. In this case, the expansion, scalar factor,
anisotropy, shear and deceleration parameters are respectively
given by
\begin{eqnarray}
\Theta=3H=\frac{\sqrt{24\pi G \rho_{0}v^{2-\gamma} +24\pi G
\rho_{ch_{0}}v^2{[A_s+{(1-A_s)}{v^{-2}}]}^{1/2}+C}}{v},\label{35}
\end{eqnarray}
\begin{eqnarray}
a_i=a_{0i} v^{1/3} \exp \left[\int {\frac{ h_idv }{\sqrt{24\pi G
\rho_{0}v^{4-\gamma} +24\pi G
\rho_{ch_{0}}v^4{[A_s+{(1-A_s)}{v^{-2}}]^{1/2}}+Cv^2}} }\right],
\quad i=1,2,3,\label{37}
\end{eqnarray}
\begin{eqnarray}
A=\frac{3h^2}{24\pi G \rho_{0}v^{2-\gamma} +24\pi G
\rho_{ch_{0}}v^2{[A_s+{(1-A_s)}{v^{-2}}]^{1/2}}+C},\label{36}
\end{eqnarray}
\begin{eqnarray}
\sigma^2=\frac{h^2}{2v^2},\label{36}
\end{eqnarray}
\begin{eqnarray}
q=2-\frac{12\pi G \rho_{0}(2-\gamma)v^{2-\gamma}+24\pi G
\rho_{ch_{0}}v^2\left[A_s+\frac{(1-A_s)}{v^{2}}\right]
^{-1/2}\left[A_s+\frac{(1-A_s)}{2v^{2}}\right]}{8\pi G
\rho_{0}v^{2-\gamma} +8\pi G
\rho_{ch_{0}}v^2{[A_s+{(1-A_s)}{v^{-2}}]^{1/2}}+C}.\label{38}
\end{eqnarray}
In figure 1 we present the dynamics of the deceleration parameter
for different values of $A_s$ and for $\gamma=\frac{4}{3}$ and
$\alpha=1$. This behavior is much dependent on the range of the
values that $A_s$ can take. For having an accelerating universe,
the value of $A_s$ should lie in the range $0\leq A_s\leq1$. In
the initial stage the evolution of the Bianchi type I brane
universe is non-inflationary, but in the late time limit the brane
universe ends in an accelerating stage.

For a better understanding of the behavior of the mean anisotropy
parameter, let us consider it as a function of the volume scale
factor
\begin{eqnarray}
A(v)=\frac{3h^2}{9\beta^{2}v^{4/3}+3\Lambda v^2+24\pi G
\rho_{0}v^{2-\gamma} +f(v)+C},\label{45}
\end{eqnarray}
where $f(v)$ is defined by equation (\ref{hypergeometry}). The
behavior of the anisotropy parameter at the initial state depends
on the values of $\alpha$ and $A_{s}$. For original Chaplygin gas
with $\alpha=1$, we obtain an accelerating universe when $0\leq
A_{s}\leq1$. From equation (\ref{45}), in the limit $v\rightarrow
0$ and taking $0\leq A_{s}\leq1$, we find that the initial state
is always anisotropic $A(v)\neq0$. The behavior of the mean
anisotropy parameter of the Bianchi type I and V geometries is
illustrated, for $\alpha=1$ and different values of $A_{s}$, in
figure 2. The behavior of this parameter shows that the universe
starts from a singular state with maximum anisotropy and ends up
in an isotropic de Sitter inflationary phase at late times. The
time variation of the shear parameter is represented for different
values of $A_s$, in figure 3.
\begin{figure}
\centerline{\begin{tabular}{ccc}
\epsfig{figure=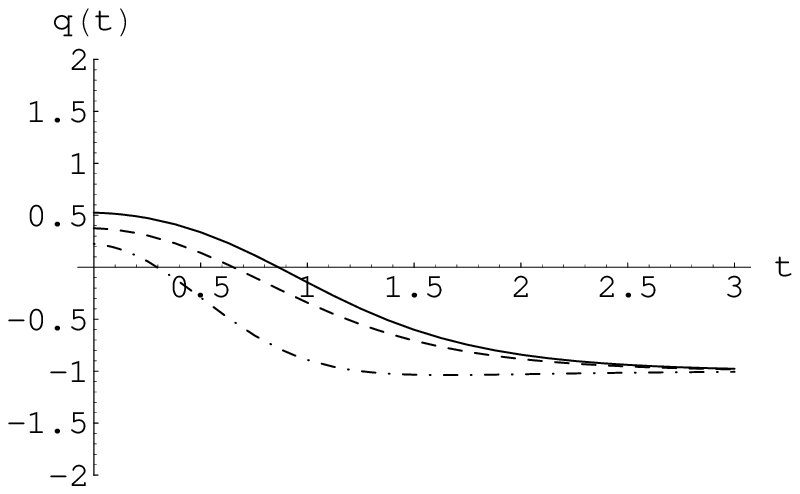,width=8cm}
\epsfig{figure=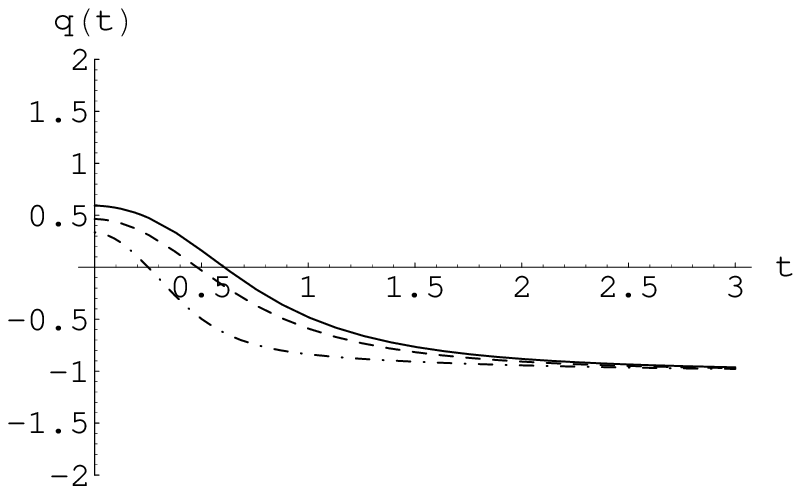,width=8cm}
\end{tabular} } \caption{\footnotesize  Left, deceleration parameter $q$
for the geometrical Chaplygin gas filled Bianchi type I brane
universe as a function of time and right, the same parameter in
the Bianchi type V brane universe for $\gamma=4/3$, $\alpha=1$ and
$A_{s}=0.3$ (solid line), $A_{s}=0.5$ (dashed line), $A_{s}=0.8$
(dot-dashed line) with $\Lambda=0$.}\label{fig0}
\end{figure}
\begin{figure}
\centerline{\begin{tabular}{ccc}
\epsfig{figure=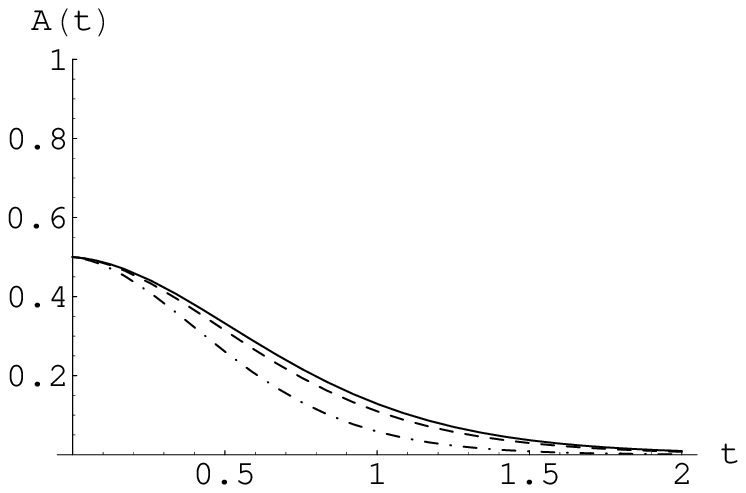,width=8cm}\epsfig{figure=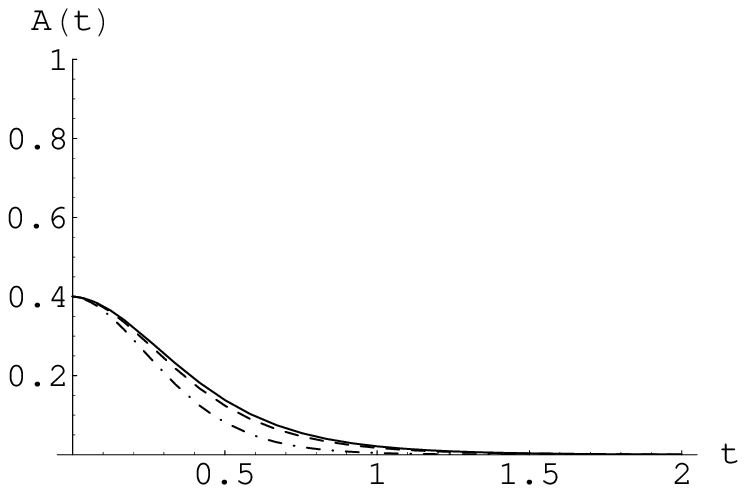,width=8cm}
\end{tabular} } \caption{\footnotesize  Left, anisotropy parameter $A$
for the geometrical Chaplygin gas filled Bianchi type I brane
universe as a function of time and right, the same parameter in
the Bianchi type V brane universe for $\gamma=4/3$, $\alpha=1$ and
$A_{s}=0.3$ (solid line), $A_{s}=0.5$ (dashed line), $A_{s}=0.8$
(dot-dashed line) with $\Lambda=0$.}\label{fig1}
\end{figure}
\begin{figure}
\centerline{\begin{tabular}{ccc}
\epsfig{figure=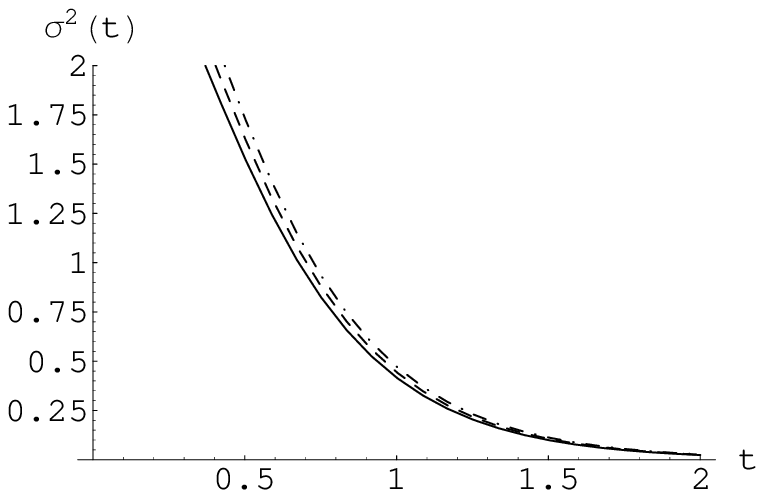,width=8cm}\epsfig{figure=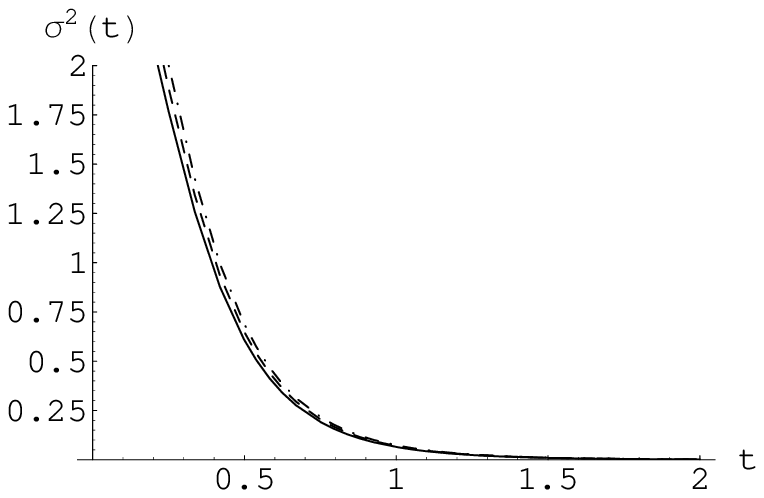,width=8cm}
\end{tabular} } \caption{\footnotesize  Left, shear scalar $\sigma^2$
for the geometrical Chaplygin gas filled Bianchi type I brane
universe as a function of time and right, the same parameter in
the Bianchi type V brane universe for $\gamma=4/3$, $\alpha=1$ and
$A_{s}=0.3 $ (solid line), $A_{s}=0.5$ (dashed line), $A_{s}=0.8$
(dot-dashed line) with $\Lambda=0$.}\label{fig2}
\end{figure}

It is worth mentioning that although most of the works on GCG
cosmology have assumed a value for $\alpha$ compatible with
$0<\alpha\leq1$, it has been shown that the type-Ia Supernovae
data favors $\alpha>1$ \cite{Bertolami1,Bertolami2,Bento}. The
best fitted values suggested for $\alpha>1$ are $\alpha=3.75$ and
$A_{s}=0.936$ \cite{Bertolami1}. In figure 4 we have plotted the
deceleration and anisotropy parameters of the Bianchi type I
geometry for $\alpha=3.75$ and different values of $A_{s}$. This
behavior shows that for $\alpha>1$ with $0\leq A_{s}\leq1$, the
geometrical model presented in this work is in agreement with
observational data.
\begin{figure}
\centerline{\begin{tabular}{ccc}
\epsfig{figure=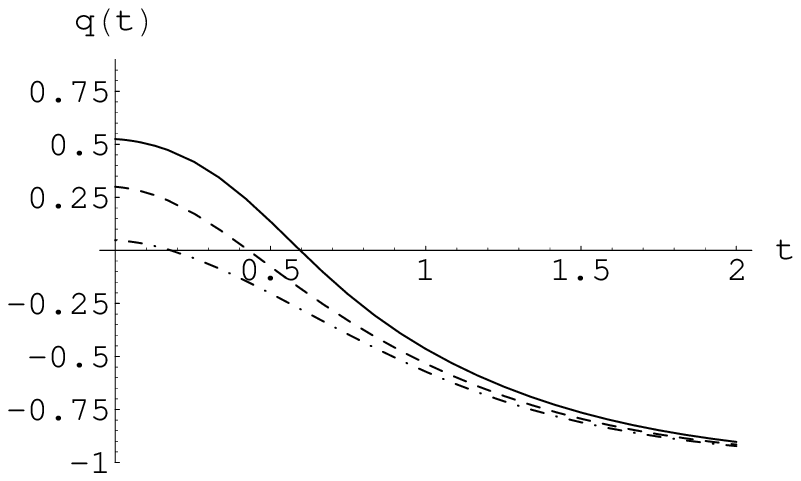,width=8cm}\epsfig{figure=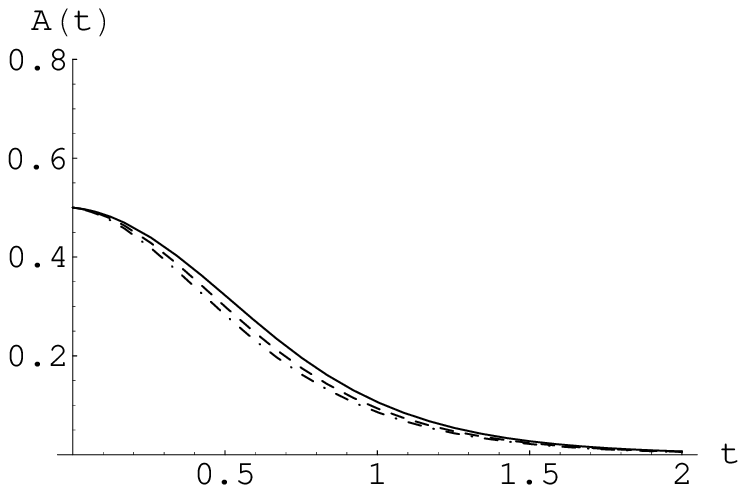,width=8cm}
\end{tabular} } \caption{\footnotesize  Left, deceleration
parameter $q$ of the Bianchi type I brane universe as a function
of time and right, anisotropy parameter $A$ of the Bianchi type I
brane universe as a function of time  for $\gamma=4/3$,
$\alpha=3.75$ and $A_{s}=0.3$ (solid line), $A_{s}=0.6$ (dashed
line), $A_{s}=0.936$ (dot-dashed line) with
$\Lambda=0$.}\label{fig2}
\end{figure}
\section{Conclusions}
In this paper we have shown that dark energy may be considered as
a consequence of the extrinsic curvature in a brane-world scenario
and extended the predictions of the geometrical matter in the more
general case when the relation between $p$ and $\rho$ is not
linear \cite{gas}. In this work, we have studied the Bianchi type
I and V geometries, seen as a brane-world embedded in a
five-dimensional bulk of constant curvature, without $Z_2$
symmetry or any form of junction condition. We have shown that the
geometrical Chaplygin gas may be used to account for the
accelerated expansion of an anisotropic universe. We have also
obtained the general solutions in an exact parametric form for
both Bianchi type I and V geometries and studied the behavior of
the observationally important parameters.

The study of anisotropic homogeneous brane-world cosmological
models has shown that in the framework of the RS models, brane
universes are born into an isotropic state \cite{Mak}, whereas in
the model presented here the universe starts as a singular state
with maximum anisotropy and reaches an isotropic state in the late
time limit, in agreement with the standard $4D$ cosmology.


\begin{thebibliography}{99}

\bibitem{3} A. Kamenshchik, U. Moschella and V. Pasquier, {\it
Phys. Lett.} B {\bf 511} 265 (2001).

\bibitem{4} M. C. Bento, O. Bertolami and A. A. Sen, {\it
Phys. Rev.} D {\bf 66} 043507 (2002).

\bibitem{5} N. Bilic, G. B. Tupper and  R.D. Viollier,
{\it Phys. Lett.} B {\bf 535} 17 (2002).

\bibitem{Kame} A. Kamenshchik, U. Moschella and V. Pasquier, {\it Phys.
Lett.} B {\bf 487} 7 (2000).

\bibitem{Kama} S. K. Kama, {\it Phys.
Lett.} B {\bf 424} 39 (1998).

\bibitem{6} J. C. Fabris, S. V. B. Goncalves and P. E. de Souza, {\it Gen. Rel. Grav.}
{\bf 34} 2111 (2002),\\ M. Bordemann and J. Hoppe,  {\it Phys.
Lett.} B {\bf 317} 315 (1993),\\ N. Ogawa, {\it Phys. Rev.} D {\bf
62} 085023 (2000).

\bibitem{9} M. C. Bento, O. Bertolami and A. A. Sen, {\it Phys. Lett.} B {\bf 575} 172 (2003).

\bibitem{SNe} M. Makler, S. Q. de Oliveira and I. Waga, {\it Phys.
Lett. } B {\bf 555} 1 (2003),\\ Y. Gong and C. K. Duan, {\it
Class. Quant. Grav.} {\bf 21} 3655 (2004).

\bibitem{CMB} M. C. Bento, O. Bertolami and A. A. Sen, {\it Phys. Lett.} B {\bf 575} 172
(2003),\\ L. Amendola, F. Finelli, C. Burigana and D. Carturan,
{\it JCAP.} {\bf 0307} 005 (2003),\\ M. C. Bento, O. Bertolami and
A. A. Sen, {\it Phys. Rev.} D {\bf 67} 063003 (2003).

\bibitem{other} R. Bean, O. Dore, {\it Phys. Rev. } D {\bf 68} 023515
(2003),\\ Z. H. Zhu, {\it Astron. Astrophys. } {\bf 423} 421
(2004).

\bibitem{Mak} M. K. Mak and
 T. Harko, {\it Phys. Rev. } D {\bf 71} 104022 (2005).

\bibitem{c1} L. P. Chimento and R. Lazkoz, {\it Phys. Lett. }
B {\bf 615} 146 (2005),\\ T. Barreiro, A. A. Sen, {\it Phys. Rev.
} D {\bf 70} 124013 (2004),\\ P. F. Gonzalez-Diaz, {\it Phys.
Lett. } B {\bf 562} 1 (2003),\\ L. P. Chimento, {\it Phys. Rev. }
D {\bf 69} 123517 (2004),\\ G. M. Kremer, {\it Phys. Rev.} D {\bf
68} 123507 (2003),\\C. S. J. Pun, L. A. Gergely, M. K. Mak, G. M.
Szabo and T. Harko, {\it Phys. Rev.} D {\bf 77} 063528 (2008).

\bibitem{C1} P. Pedram and S. Jalalzadeh, {\it Phys. Lett.} B {\bf 659} 6 (2008),\\
U. Debnath, A. Banerjee and S. Chakraborty, {\it Class. Quant.
 Grav. } {\bf 21} 5609 (2004),\\ M. C. Bento, O. Bertolami and A. A. Sen, {\it Phys. Rev.} D {\bf
70} 083519 (2004),\\ R. R. R. Reis, I. Waga, M. O. Calvao and S.
E. Joras, {\it Phys. Rev.} D {\bf 68} 061302 (2003),\\ T.
Multamaki, M. Manera and E. Gaztanaga, {\it Phys. Rev.} D {\bf 69}
023004 (2004),\\ P. P. Avelino, L. M. G. Beca, J. P. M. de
Carvalho and C. J. A. P. Martins, {\it JCAP} {\bf 0309} 002
(2003).

\bibitem{Nima} N. Arkani-Hamed, S. Dimopoulos, and G. Dvali, {\it Phys.
Lett.} B {\bf 429} 263 (1998),\\ I. Antoniadis, N. Arkani-Hamed,
S. Dimopoulos, and G. Dvali, {\it Phys. Lett.} B {\bf 436} 257
(1998).

\bibitem{Randall} L. Randall and R. Sundrum, {\it Phys. Rev. Lett.}
{\bf 83}  3370 (1999),\\ L. Randall and R. Sundrum,   {\it Phys.
Rev. Lett.} {\bf 83}  4690 (1999).

\bibitem{Dvali} G. Dvali, G. Gabadadze, and M. Porrati, {\it Phys. Lett.} B {\bf 485} 208
(2000),
\\ G. Dvali and G. Gabadadze, {\it Phys. Rev.} D {\bf 63} 065007
(2001).

\bibitem{israel} W. Israel, {\it Nuovo Cimento} B {\bf 44}
1 (1966).

\bibitem{Battye} R. A. Battye and B. Carter, {\it Phys. Lett.} B {\bf 509} 331
(2001).

\bibitem{Cline} J. M. Cline, C. Grojean and G. Servant, {\it Phys.
Rev. Lett.} {\bf 83} 4245 (1999),\\ T. Shiromizu, K. Maeda and M.
Sasaki, {\it Phys. Rev.} D {\bf 62} 024012 (2000).

\bibitem{Binetruy} P. Binetruy, C. Deffayet and D. Langlois, {\it Nucl.
Phys.} B {\bf 565} 269 (2000),\\ P. Binetruy, C. Deffayet, U.
Ellwanger and D. Langlois, {\it Phys. Lett.} B {\bf 477} 285
(2000).

\bibitem{Maia} M. D. Maia, E. M. Monte and J. M. F. Maia, {\it Phys.
Lett.} B {\bf 585} 11 (2004).

\bibitem{maia} M. D. Maia, E. M. Monte, J. M. F. Maia and J. S. Alcaniz,
{\it Class. Quant. Grav.} {\bf 22} 1623 (2005).

\bibitem{Rubakov} V. A. Rubakov and M. E. Shaposhnikov, {\it Phys. Lett.} B
{\bf 125} 136 (1983).

\bibitem{shahram} S. Jalazadeh and H. R. Sepangi, {\it Class. Quant. Grav.}
{\bf 22} 2035 (2005).

\bibitem{fard} M. Heydari-Fard, M. Shirazi, S. Jalalzadeh and H. R.
Sepangi, {\it Phys. Lett.} B {\bf 640} 1 (2006).

\bibitem{gas} M. Heydari-Fard and H. R. Sepangi, {\it Phys. Rev.} D {\bf 76}
104009 (2007).

\bibitem{Nash} J. Nash, {\it Ann. Maths.} {\bf 63} 20 (1956).

\bibitem{Book} L. P. Eisenhart  1966 {\it Riemannian
Geometry}, Princeton University Press, Princeton NJ (1966).

\bibitem{Chen} C. M. Chen, T. Harko and M. K. Mak, {\it Phys.
Rev.} D {\bf 64} 044013 (2001).



\bibitem{Bertolami1} O. Bertolami, A. A. Sen, S. Sen and P. T. Silva, {\it Mon. Not. Roy. Astron. Soc.} {\bf 353} 329
(2004).



\bibitem{Bertolami2} M. C. Bento, O. Bertolami, N. M. C. Santos and A. A. Sen, {\it
Phys. Rev.} D {\bf 71}  063501 (2005).



\bibitem{Bento} M. C. Bento, O. Bertolami, M. J. Rebouças and P. T. Silva, {\it Phys. Rev.} D {\bf 73}  043504
(2006),\\ A. A. Sen, R. J. Scherrer, {\it Phys. Rev.} D {\bf 72}
063511 (2005),\\ Y. Gong, {\it JCAP} {\bf 0503} 007 (2005).

\end{thebibliography}
\end{document}